\documentstyle[sprocl,epsfig]{article}

\def\be{\begin{equation}}
\def\ee{\end{equation}}
\def\bea{\begin{eqnarray}}
\def\eea{\end{eqnarray}}

\begin{document}

\title{SHAPE PHASE TRANSITIONS AND RANDOM INTERACTIONS}

\author{Roelof BIJKER}

\address{ICN-UNAM, AP 70-543, 04510 M\'exico, DF, M\'exico}

\maketitle

\abstracts{
The phenomenom of emerging regular spectral features from random 
interactions is addressed in the context of the interacting boson model. 
A mean-field analysis links different regions of the 
parameter space with definite geometric shapes. The results 
provide a clear and transparent interpretation of the high degree 
of order that has been observed before in numerical studies.}

\section{Introduction}

Recent shell model calculations for even-even nuclei in the $sd$ shell and 
the $pf$ shell showed, despite the random nature of the two-body matrix 
elements, a remarkable statistical preference for ground states with angular 
momentum $L=0$ \cite{JBD}. A similar dominance of $L=0$ ground states was 
found in an analysis of the Interacting Boson Model (IBM) with random 
interactions \cite{BF1}. In addition, in the IBM there is strong evidence 
for both vibrational and rotational band structures. According to the 
conventional ideas in the field, the occurrence of regular spectral features 
is due to a very specific form of the interactions. The studies with random 
interactions show that the class of Hamiltonians that lead to these ordered 
patterns is much larger than is usually thought. 

The basic ingredients of the numerical simulations, both for the 
nuclear shell model and for the IBM, are the structure of the 
model space, the ensemble of random Hamiltonians, the order of the 
interactions (one- and two-body), and the global symmetries, i.e. 
time-reversal, hermiticity and rotation and reflection symmetry. 
The latter three symmetries of the Hamiltonian cannot be modified, 
since we are studying many-body systems whose eigenstates have real 
energies and good angular momentum and parity. It was found that the 
observed spectral order is a rather robust property which does 
not depend on the specific  choice of the (two-body) ensemble of 
random interactions \cite{JBD,BFP1,JBDT,DD}, the time-reversal 
symmetry \cite{BFP1}, or the restriction of the Hamiltonian to one- 
and two-body interactions \cite{BF2}. This suggests that 
that an explanation of the origin of the observed regular features 
has to be sought in the many-body dynamics of the model space 
and/or the general statistical properties of random interactions. 

The purpose of this contribution is to investigate the distribution of 
ground state angular momenta for the IBM in a Hartree-Bose mean-field 
analysis \cite{BF3}. 

\section{Phase transitions}

The IBM describes low-lying collective excitations in nuclei in terms 
of a system of $N$ interacting quadrupole ($d^{\dagger}$) and monopole 
($s^{\dagger}$) bosons \cite{IBM}. The IBM Hamiltonian spans a wide range 
of collective features which includes vibrational, rotational and $\gamma$ 
unstable nuclei. The connection with potential energy surfaces, geometric 
shapes and phase transitions can be studied by means of Hartree-Bose 
mean-field methods \cite{cs,duke} in which the trial wave function is written 
as a coherent state. For one- and two-body interactions the coherent state 
can be expressed in terms of an axially symmetric condensate 
\bea
\left| \, N,\alpha \, \right> \;=\; \frac{1}{\sqrt{N!}} 
\left( \cos \alpha \, s^{\dagger} + \sin \alpha \, d_0^{\dagger} 
\right)^N \, \left| \, 0 \, \right> ~, 
\label{trial}
\eea
with $-\pi/2 < \alpha \leq \pi/2$. The angle $\alpha$ is related to 
the deformation parameters in the intrinsic frame, $\beta$ and $\gamma$ 
\cite{cs}. First we investigate the properties of some schematic 
Hamiltonians that have been used in the study of shape phase 
transitions.   

\subsection{The $U(5)$-$SO(6)$ case}
 
The transition from vibrational to $\gamma$ unstable nuclei can be described 
by the Hamiltonian 
\bea
H \;=\; \frac{\cos \chi}{N} \, d^{\dagger} \cdot \tilde{d} +
\frac{\sin \chi}{4N(N-1)} \,
( s^{\dagger} s^{\dagger} - d^{\dagger} \cdot d^{\dagger} ) \,
( \tilde{s} \tilde{s} - \tilde{d} \cdot \tilde{d} ) ~, 
\label{hcase1}
\eea
which exhibits a second order phase transition at $\chi_c=\pi/4$ \cite{cs}. 
For the present application, we extend the range of the angle $\chi$ to that 
of a full period $-\pi/2 < \chi \leq 3\pi/2$, so that all possible 
combinations of attractive and repulsive interactions are included. 
The potential energy surface is given by the expectation value of $H$ 
in the coherent state 
\bea
E(\alpha) \;=\; \cos \chi \, \sin^2 \alpha 
+ \frac{1}{4} \sin \chi \, \cos^2 2\alpha ~.
\eea
The equilibrium configurations are characterized by the value of 
$\alpha=\alpha_0$ for which the energy surface has its minimum. They 
can be divided into three different classes or shape phases 
\bea
\begin{array}{rclcrcl}
\alpha_0 &=& 0 & \hspace{1cm}  & -\pi/2 < &\chi& \leq \pi/4 \\
\cos 2\alpha_0 &=& \cot \chi & &  \pi/4 \leq &\chi& \leq 3\pi/4 \\
\alpha_0 &=& \pi/2 & & 3\pi/4 \leq &\chi& \leq 3\pi/2
\end{array}
\eea
which correspond to an $s$-boson or spherical condensate, a deformed 
condensate, and a $d$-boson condensate, respectively. The phase transitions 
at the critical points $\chi_{\rm c}=\pi/4$ and $3\pi/4$ are of second 
order, whereas the one at $3\pi/2$ is of first order. 

The angular momentum of the ground state can be obtained from the rotational 
structure of the equilibrium configuration, in combination with the 
Thouless-Valatin formula for the corresponding moments 
of inertia \cite{duke}. \\
$\bullet$ For $\alpha_0=0$ the equilibrium configuration has spherical 
symmetry, and hence can only have $L=0$. \\
$\bullet$ For $0 < \alpha_0 < \pi/2$ the condensate is deformed. 
The ordering of the rotational energy levels $L=0,2,\ldots,2N$ 
\bea
E_{\rm rot} \;=\; \frac{1}{2{\cal I}_3} L(L+1) ~,
\eea
is determined by the sign of the moment of inertia 
\bea
{\cal I}_3 \;=\; \frac{3N(\sin \chi - \cos \chi)}{\sin \chi \cos \chi} ~. 
\eea
For $\pi/4 \leq \chi \leq \pi/2$ the moment of inertia ${\cal I}_3$ is 
positive and hence the ground state has angular momentum $L=0$, whereas for 
for $\pi/2 \leq \chi \leq 3\pi/4$ it is negative corresponding to a 
ground state with $L=2N$. \\
$\bullet$ For $\alpha_0=\pi/2$ we find a condensate of $N$ 
quadrupole or $d$-bosons, which corresponds to a quadrupole oscillator 
with $N$ quanta. Its rotational structure is characterized by the labels 
$\tau$, $n_{\Delta}$ and $L$. The boson seniority $\tau$ is given by 
$\tau=3n_{\Delta}+\lambda=N,N-2,\ldots,1$ or $0$ for $N$ odd or even, 
and the values of the angular momenta are 
$L=\lambda,\lambda+1,\ldots,2\lambda-2,2\lambda$ \cite{IBM}. 
In general, the rotational excitation energies depend on two moments of 
inertia 
\bea
E_{\rm rot} \;=\; \frac{1}{2{\cal I}_5} \tau(\tau+3) 
+ \frac{1}{2{\cal I}_3} L(L+1) ~. 
\label{erot}
\eea
For the special case of the Hamiltonian of Eq.~(\ref{hcase1}) only the 
first term is present
\bea
{\cal I}_5 \;=\; -\frac{2N}{\sin \chi} ~.
\eea
For $3\pi/4 \leq \chi \leq \pi$ the moment of inertia ${\cal I}_5$ is 
negative and the ground state has $\tau=N$, whereas for 
$\pi \leq \chi \leq 3\pi/2$ it is positive and the ground state has 
$\tau=0$ ($L=0$) for $N$ even, and $\tau=1$ ($L=2$) for $N$ odd. 

\begin{figure}
\centerline{\hbox{\epsfig{figure=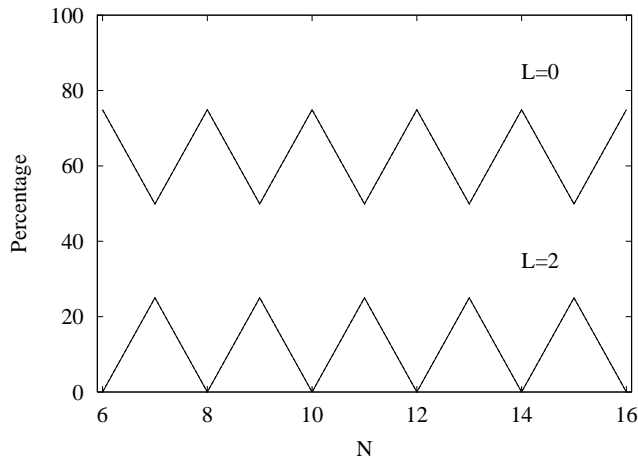,width=0.75\linewidth} }}
\caption[]{Percentages of ground states with $L=0$ and $L=2$ for the 
schematic IBM Hamiltonian of Eq.~(\protect\ref{hcase1}) with 
$-\pi/2 < \chi \leq 3\pi/2$ calculated exactly (solid lines) and in 
mean-field approximation (dashed lines).}
\label{trans1}
\end{figure}

In Fig.~\ref{trans1} we compare the percentages of ground states with $L=0$ 
and $L=2$ as a function of $N$ obtained exactly (solid lines) and in the 
mean-field analysis (dashed lines). The results  were obtained by assuming 
a constant probability distribution for $\chi$ on the interval 
$-\pi/2 < \chi \leq 3\pi/2$. We have added a small attractive 
$\vec{L} \cdot \vec{L}$ interaction to remove the 
degeneracy of the ground state for the $\tau=N$ solution. There is a perfect  
agreement for all values of $N$. The ground state is most likely to have 
angular momentum $L=0$: in $75 \%$ of the cases for $N$ even and in $50 \%$ 
for $N$ odd. In $25 \%$ of the cases, the ground state has the maximum value 
of the angular momentum $L=2N$. The only other value that occurs is $L=2$ in 
$25 \%$ of the cases for $N$ odd. The oscillation in the $L=0$ and $L=2$ 
percentages is due to the contribution of the $d$-boson condensate. The sum 
of the $L=0$ and $L=2$ percentages is constant ($75 \%$) and does not depend 
on $N$. 

\subsection{The $U(5)$-$SU(3)$ case} 

\begin{figure}
\centerline{\hbox{\epsfig{figure=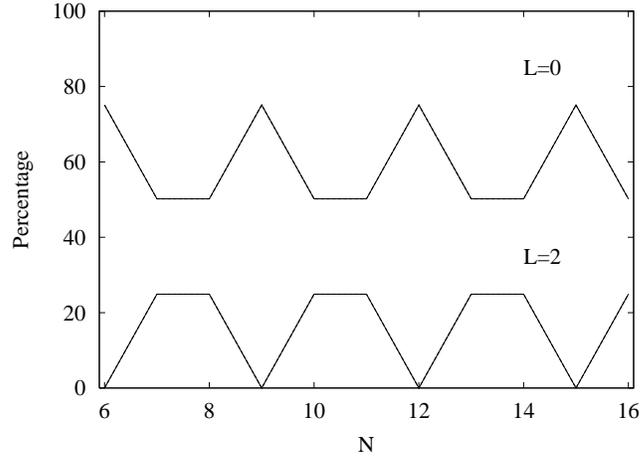,width=0.75\linewidth} }}
\caption[]{As Fig.~\protect\ref{trans1}, but for the schematic IBM 
Hamiltonian of Eq.~(\protect\ref{hcase2}).}
\label{trans2}
\end{figure}

A second transitional region of interest is the one between 
vibrational and rotational nuclei. In the IBM, it can be described 
schematically by 
\bea
H_{\pm} &=& \cos \chi \, d^{\dagger} \cdot \tilde{d} +
\frac{\sin \chi}{N-1} \left[
( 2 \, s^{\dagger} \cdot s^{\dagger} - d^{\dagger} \cdot d^{\dagger} ) \,
( 2 \, \tilde{s} \cdot \tilde{s} - \tilde{d} \cdot \tilde{d} ) \right. 
\nonumber\\
&& \left. + ( 2 \, s^{\dagger} \times d^{\dagger} \pm \sqrt{7} \, 
d^{\dagger} \times d^{\dagger} )^{(2)} \cdot ( 2 \, \tilde{d} \times 
\tilde{s} \pm \sqrt{7} \, \tilde{d} \times \tilde{d} )^{(2)}  
\right] ~. 
\label{hcase2}
\eea
In the physical region $0 \leq \chi \leq \pi/2$, $H_{\pm}$ exhibits a first 
order phase transition at $\chi_c=\arctan 1/9$ \cite{cs}. As before, here 
we consider the interval $-\pi/2 < \chi \leq 3\pi/2$. 
The results for the distribution of ground state angular momenta are 
presented in Fig.~\ref{trans2}. For $N=3k$ the ground state has $L=0$ 
in $75 \%$ of the cases and $L=2N$ in the remaining $25 \%$. For $N=3k+1$ 
and $N=3k+2$ the ground state angular momentum is either $L=0$ ($50 \%$), 
$L=2$ ($25 \%$) or $L=2N$ ($25 \%$). The variation in the $L=0$ and $L=2$ 
percentages is due to the contribution of the $d$-boson condensate, 
whereas the sum of the two is constant ($75 \%$). 

\subsection{The $SU(3)$-$SO(6)$ case} 

\begin{figure}
\centerline{\hbox{\epsfig{figure=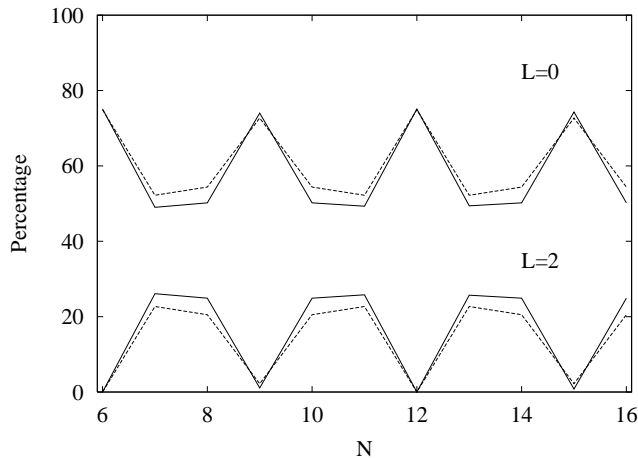,width=0.75\linewidth} }}
\caption[]{As Fig.~\protect\ref{trans1}, but for the schematic IBM 
Hamiltonian of Eq.~(\protect\ref{hcase3}).}
\label{trans3}
\end{figure}

The transitional region between rotational and $\gamma$ unstable nuclei 
described by the Hamiltonian 
\bea
H_{\pm} &=& \frac{\cos \chi}{4(N-1)} \,
( s^{\dagger} \cdot s^{\dagger} - d^{\dagger} \cdot d^{\dagger} ) \,
( \tilde{s} \cdot \tilde{s} - \tilde{d} \cdot \tilde{d} ) 
\nonumber\\
&& + \frac{\sin \chi}{N-1} \left[
( 2 \, s^{\dagger} \cdot s^{\dagger} - d^{\dagger} \cdot d^{\dagger} ) \,
( 2 \, \tilde{s} \cdot \tilde{s} - \tilde{d} \cdot \tilde{d} ) \right. 
\nonumber\\
&& \left. + ( 2 \, s^{\dagger} \times d^{\dagger} \pm \sqrt{7} \, 
d^{\dagger} \times d^{\dagger} )^{(2)} \cdot ( 2 \, \tilde{d} \times 
\tilde{s} \pm \sqrt{7} \, \tilde{d} \times \tilde{d} )^{(2)} 
\right] ~,
\label{hcase3}
\eea
does not show a phase transition in the physical region 
$0 \leq \chi \leq \pi/2$ \cite{cs}. Fig.~\ref{trans3} shows that the 
distribution of the ground state angular momenta is very similar to the 
previous case. 

\section{Random interactions}

Finally, we apply the mean-field analysis to the general one- and 
two-body IBM Hamiltonian 
\bea
H \;=\; \frac{1}{N} \left[ H_1 + \frac{1}{N-1} H_2 \right] ~, 
\label{h12}
\eea
in which the nine parameters of this Hamiltonian are taken as independent 
random numbers on a Gaussian distribution with zero mean and width $\sigma$. 
The distribution of geometric shapes for this ensemble of Hamiltonians 
is determined by the distribution of equilibrium configurations 
of the corresponding potential energy surfaces 
\bea
E(\alpha) \;=\; a_4 \, \sin^4 \alpha + 
a_3 \, \sin^3 \alpha \cos \alpha + a_2 \, \sin^2 \alpha + a_0 ~. 
\label{surface}
\eea
The coefficients $a_i$ are linear combinations of the Hamiltonian parameters. 
The spectral properties of each Hamiltonian of the ensemble of random 
one- and two-body interactions are analyzed by exact numerical 
diagonalization \cite{BF1} and by mean-field analysis \cite{BF3}.   

\begin{figure}
\centerline{\hbox{\epsfig{figure=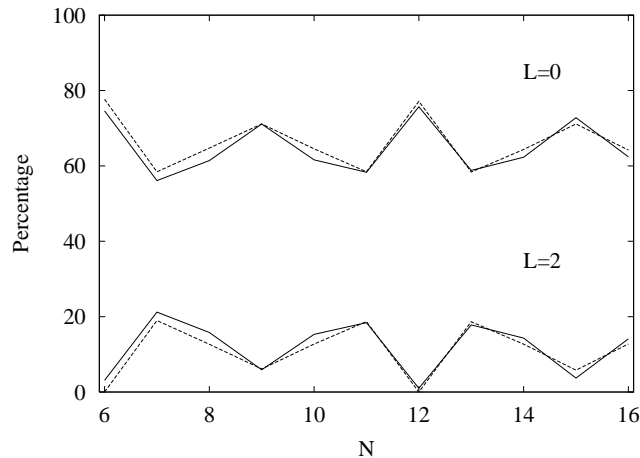,width=0.75\linewidth} }}
\caption[]{As Fig.~\protect\ref{trans1}, but for the random IBM 
Hamiltonian of Eq.~(\protect\ref{h12}).}
\label{ibmgs}
\end{figure}

In Fig.~\ref{ibmgs} we compare the percentages of $L=0$ and $L=2$ ground 
states obtained exactly (solid lines) and in the mean-field analysis (dashed 
lines). There is a dominance of ground states with $L=0$ for $\sim 63-77 \%$ 
of the cases. 
For $N=3k$ we see an enhancement for $L=0$ and a corresponding decrease for 
$L=2$. Also in this case, the equilibrium configurations can be divided into 
three different classes: an $s$-boson or spherical condensate, a deformed 
condensate, and a $d$-boson condensate. For the spherical and deformed 
solutions the ground state has $L=0$ ($\sim 63 \%$) or $L=2N$ ($\sim 13 \%$). 
The analysis of the $d$-boson condensate is a bit more complicated due to 
the presence of two moments of inertia, ${\cal I}_5$ and ${\cal I}_3$. There 
is a constant contribution to $L=2N$ ground states ($\sim 10 \%$), whereas 
the $L=0$ and $L=2$ percentages show oscillations with $N$ \cite{BF3}. 
Just as for the schematic Hamiltonians, the sum of the $L=0$ and $L=2$ 
percentages is constant and independent of $N$. 

\section{Summary and conclusions}

In this contribution, we have investigated the origin of the regular features 
that have been observed in numerical studies of the IBM with random 
interactions, in particular the dominance of ground states with $L=0$. 

In a mean-field analysis, it was found that different regions of the 
parameter space can be associated with particular intrinsic vibrational 
states, which in turn correspond to definite geometric shapes: a spherical 
shape, a deformed shape or a condensate of quadrupole bosons. An analysis 
of the angular momentum content of each one of the corresponding condensates 
combined with the sign of the relevant moments of inertia, provides an 
explanation for the distribution of ground state angular momenta of 
both schematic and random forms of the IBM Hamiltonian. 

In summary, the present results show that mean-field methods provide a clear 
and transparent interpretation of the regular features that have been 
obtained before in numerical studies of the IBM with random interactions. 
The same conclusions hold for the vibron model \cite{BF4}. For the nuclear 
shell model the situation is less clear. Despite the large number of studies 
that have been carried out to explain and/or further explore the properties 
of random nuclei no definite answer is yet available \cite{review}. 

\section*{Acknowledgments}

It is a great pleasure to dedicate this contribution to the 60th 
birthday of Jerry P. Draayer. Congratulations, Jerry! 
This work was supported in part by CONACyT under project No. 32416-E.

\end{document}